\documentstyle[12pt]{article}
\begin{document}
\newcommand{\be}{\begin{equation}}
\newcommand{\ee}{\end{equation}}
\newcommand{\ba}{\begin{array}}
\newcommand{\ea}{\end{array}}
\newcommand{\rar}{\rightarrow}
\newcommand{\fr}{\frac}
\newcommand{\nl}{\newline}
\newcommand{\hs}{\hspace}
\newcommand{\vs}{\vspace}
\newcommand{\hst}{\hspace*}
\newcommand{\vst}{\vspace*}
\newcommand{\lb}{\label}
\newcommand{\al}{\alpha}
\newcommand{\bt}{\beta}
\newcommand{\eps}{\epsilon}
\newcommand{\veps}{\varepsilon}
\newcommand{\ld}{\lambda}
\newcommand{\sg}{\sigma}
\newcommand{\Sg}{\Sigma}
\newcommand{\bib}{\bibitem}
\newcommand{\ct}{\cite}
\newcommand{\rf}{\ref}
\newcommand{\PRL}[1]{{\em Phys. Rev. Lett.}\ {\bf #1}}
\title{The Majorana Fermions for Quantum S=1/2 Antiferromagnet\,?\vst{0.6cm}}
\author{\sc Alexander Moroz\thanks{e-mail address :
{\tt moroz@eldp.epfl.ch}}}
\date{\it Institute of Theoretical Physics\\
\it EPFL\\
\it PHB-Ecublens\\
\it CH-1015 Lausanne, Suisse}
\maketitle
\begin{center}
{\large\sc abstract}
\end{center}
\vspace{0.6cm}
Recently it has been suggested by A. M. Tsvelik that quantum S=1/2
antiferromagnet can be described by the Majorana fermions
in an irreducible way and without any constraint. In contrast to this
claim we shall show that this representation is highly reducible. It is
a direct sum  of four irreducible fundamental representations
of $su(2)$ algebra.
\thispagestyle{empty}
\baselineskip 20pt
\newpage
\noindent
When dealing with the  Heisenberg spin $S=1/2$  antiferromagnet
one usually represents the spin operators $S^a{}'s$
in terms of fermionic degrees of freedom. In the Schwinger representation
of spin operators,
\be
S^a =b^+_\al\sg^a_{\al\bt}b_\bt,
\lb{sr}
\ee
where the fermionic degrees of freedom satisfy the constraint
\be
b^+_\al b_\al =1,
\lb{sr1}
\ee
and $\sg^a{}'s$ are the Pauli matrices.
As one easily finds this representation possesses $U(1)$ gauge invariance.
However this $U(1)$ gauge symmetry is known to be strongly restricted due
to a plaquette identity. On a square lattice this identity allows only
for such $U(1)$ gauge configurations (on bonds of a lattice) which give
$Z_2$-flux plaquette configurations \ct{M}.
Recently in this preprint network  a paper by
A. M. Tsvelik \ct{T} has appeared where a new fermionic description of
a quantum $S=1/2$ antiferromagnet without constraint
has been suggested in terms of the Majorana fermions.
A proof has been announced that low-lying excitations in a spin
liquid state of $S=1/2$ antiferromagnet are $S=1$ fermions.

In notation of \ct{T} the Majorana fermions on a lattice site $r$ are
denoted by $\eta_a (r)$, $a=1,2,3$. They satisfy the Clifford algebra:
\be
\{\eta_a(r),\eta_b (r')\}_+ =\delta_{ab}\delta_{rr'}.
\ee
The spin operators are represented as bilinears in the Majorana
fermions,
\be
S^a(r) :=-\fr{i}{2}\varepsilon_{abc}\eta_b(r)\eta_c(r).
\lb{rp}
\ee
One checks that (\rf{rp}) reproduces the usual commutation relations of
the spin operators,
\be
[S^a(r),S^b(r')]=i\varepsilon^{abc}\delta_{rr'}S^c(r).
\ee
No constraint on the fermionic degrees of freedom is needed provided
the representation is irreducible.
This irreducibility was inferred in \ct{T} from
\be
S^a(r)S^a(r)=\fr{1}{4},\hs{2cm}\sum_a S^a(r)S^a(r)=\fr{3}{4}.
\lb{last}
\ee
In contrast to the Schwinger representation (\rf{sr}-\rf{sr1}) this
representation possesses $Z_2$ gauge invariance only,
\be
\eta_a(r)\rar (-1)^{q(r)}\eta_a(r),
\ee
$q(r)=\pm 1$.

At first sight this approach may seem to be appealing since apart
from missing constraint
it seems to treat directly physical $Z_2$ degrees of freedom.
However, as we shall show in a while, in contrast to the claim in \ct{T}
the representation {\em is not irreducible}.
First of all there is a {\em mismatch between the number of states} in the
fundamental representation of $su(2)$ algebra and
the representation (\rf{rp}). The representation space of the Majorana
fermions
is {\em eight-dimensional} and not two-dimensional as it should
be for the fundamental representation of $su(2)$.
It is built up by vacuum vector and
by $\eta_a{}'s$, $\eta_a\eta_b$, $a\neq b$, and $\eta_1\eta_2\eta_3$.
Moreover the vacuum is not invariant under the action of spin operators.
This is the reason why this approach cannot be extended to treat
holes, i.e., to deal with the Hubbard model in the limit $U\rar\infty$
away from half filling.

The above dimensional argument together with (\rf{last}) suggest that
the representation (\rf{rp}) is a {\em direct sum of four fundamental
representations of} $su(2)$.
One immediately finds that vector spaces generated by vectors
$\{\eta_1,\eta_2,\eta_3,\\ \eta_1\eta_2\eta_3\}$ and
$\{\eta_1\eta_2,\eta_1\eta_3,\eta_2\eta_3, 0\}$ are {\em invariant
spaces} under the action of spin operators $S^a{}'s$.
Therefore these spaces are four dimensional, one expects that they
can be further reduced.
This can be confirmed as follows. Let us take one of them, say
$\{\eta_1,\eta_2,\eta_3,\eta_1\eta_2\eta_3\}_{span}$.
In the above  basis,
\be
S^1=\left(
\ba{cccc}
0&0&0&i/4\\
0&0&-i/2&0\\
0&i/2&0&0\\
-i&0&0&0
\ea\right),\hs{0.7cm}
S^2=\left(\ba{cccc}
0&0&i/2&0\\
0&0&0&i/4\\
-i/2&0&0&0\\
0&-i&0&0
\ea\right),
\ee
\be
S^3=\left(
\ba{cccc}
0&i/2&0&0\\
-i/2&0&0&0\\
0&0&0&-i/4\\
0&0&i&0
\ea\right).
\ee
To find invariant subspaces we shall look for
the spectrum of $S^3$. Since $\{\eta_1,\eta_2\}
_{span}$
and $\{\eta_3,\eta_1\eta_2\eta_3\}_{span}$ are invariant subspaces
under the action of $S^3$, one can  diagonalize $S^3$ independently
in any of them. One finds eigenvectors
$e_+=i\eta_1 +\eta_2$ and $e_- =\eta_1+i\eta_2$ with respective
eigenvalues $\ld_{\pm}=\pm 1/2$ in the first invariant subspace, and
$\bar{e}_+=(1/2)\eta_3 +i\eta_1\eta_2\eta_3$ and $\bar{e}_- =
(i/2)\eta_3+\eta_1\eta_2\eta_3$ with respective eigenvalues
$\bar{\ld}_{\pm}=\pm 1/2$ in the second invariant subspace.
Eventually, one checks that the spaces $\{e_+,\bar{e}_-\}_{span}$
and $\{\bar{e}_+,e_-\}_{span}$ are {\em invariant
irreducible spaces under the action of the spin operators} $S^a{}'s$.
Thus they provide fundamental {\em two dimensional} representation of
$su(2)$. The original Majorana representation is the direct sum
of them. This is immediately seen from the block diagonal form of the spin
operators in the invariant subspace  $\{e_+,\bar{e}_-,\bar{e}_+,e_-\}_{span}$
($\equiv\{\eta_1,\eta_2,\eta_3,\eta_1\eta_2\eta_3\}_{span}$),
\be
S^1=\left(
\ba{cccc}
0&1/4&0&0\\
1&0&0&0\\
0&0&0&-1\\
0&0&-1/4&0
\ea\right),\hs{0.7cm}
S^2=\left(\ba{cccc}
0&i/4&0&0\\
-i&0&0&0\\
0&0&0&-i\\
0&0&i/4&0
\ea\right),
\ee
\be
S^3=\left(
\ba{cccc}
1/2&0&0&0\\
0&-1/2&0&0\\
0&0&1/2&0\\
0&0&0&-1/2
\ea\right).
\ee
Therefore the Majorana representation provides nothing but the fermionic
description of a
``four-flavor" Heisenberg spin $S=1/2$ antiferromagnet.

If one confines oneselves to
a particular invariant subspace, say $\{e_+,\bar{e}_-\}_{span}$,
one finds that the basis vectors $e_+$ and $\bar{e}_-$ considered as
operators satisfy the relations
$\{e_+,\bar{e}_-\}_+=[e_+,\bar{e}_-]_-=e_+\bar{e}_-$,
since $\bar{e}_- e_+ =0$. Thus in a given invariant subspace
one can work without constraints provided one orders all
operators on a given site before calculating expectation values
with $\bar{e}_-$ to the left of $e_+$.
This is equivalent to the Gutzwiller projection. Therefore there is no
additional advantage by using the Majorana representation (\rf{rp})
over the usual Schwinger representation of spin operators
(\rf{sr}-\rf{sr1}).\vst{0.8cm}\nl

\end{document}